\documentclass[twocolumn,prl,aps,superscriptaddress,showpacs]{revtex4}

\usepackage{bm}
\usepackage{graphicx}
\usepackage{color}

\begin{document}

\title{Ising incommensurate spin resonance of CeCoIn$_{5}$: \\ a dynamical precursor of the Q-phase.}

\author{S. Raymond}
\affiliation{Universit\'e Grenoble Alpes, INAC-SPSMS, F-38000 Grenoble, France}
\affiliation{CEA, INAC-SPSMS, F-38000 Grenoble, France}
\author{G. Lapertot}
\affiliation{Universit\'e Grenoble Alpes, INAC-SPSMS, F-38000 Grenoble, France}
\affiliation{CEA, INAC-SPSMS, F-38000 Grenoble, France}
\date{\today}

\begin{abstract}
It is shown by detailed inelastic neutron scattering experiments that the gapped collective magnetic excitation of the unconventional superconductor CeCoIn$_{5}$, the spin resonance mode, is incommensurate and that the corresponding fluctuations are of Ising nature. The incommensurate peak position of these fluctuations corresponds to the propagation vector of the adjacent field induced static magnetic ordered phase, the so-called Q-phase. 
Furthermore, the direction of the magnetic moment fluctuations is also the direction of the ordered magnetic moments of the Q-phase.
Hence the resonance mode and the Q-phase share the same symmetry and this strongly supports a scenario where the static order is realized by a condensation of the magnetic excitation. 

\end{abstract}

\maketitle

Unconventional superconductivity is reported for compounds belonging to various families of materials spanning $d$ and $f$-electron physics : cuprates, iron based materials and heavy fermion systems.
Despite this diversity, phenomenological similarities emerge indicating a possible common underlying physics \cite{Uemura,Scalapino}.
Firstly, unconventional superconductivity often occurs on the verge of magnetic ordering. Secondly, a universal feedback  of unconventional superconductivity manifests on the magnetic excitation spectrum, measured by inelastic neutron scattering, via the appearance of a new well-defined mode in the superconducting phase: the resonance peak \cite{Yu}. These points motivate a continuous theoretical and experimental effort in the parallel investigations of such systems, in particular in view of the possible common mechanism of spin fluctuation mediated superconductivity. In this context, the study of the relationship between the resonance mode and the adjacent long-range magnetically ordered phase is of first importance and can give insight into the interplay between magnetism and superconductivity. The heavy fermion superconductor CeCoIn$_{5}$ provides a unique opportunity for such an investigation on an $f$-electron system since both the dynamical resonance mode associated with superconductivity and long range magnetic ordering are reported in this system for zero and respectively finite applied magnetic field. 

CeCoIn$_{5}$ has the highest superconducting transition temperature, $T_{c}$= 2.3 K, among Ce-based heavy fermion systems \cite{Petrovic}. It crystallizes in the tetragonal space group P4/mmm and it is established that the superconducting gap symmetry is of the singlet $d_{x2-y2}$ type \cite{Allan,Zhou}. In this compound, the spin resonance is observed at the antiferromagnetic wave-vector $\bf{Q_{AF}}$=(0.5, 0.5, 0.5) for an energy of 0.6 meV ($\approx$ 7 K) that scales approximately with 3.$k_{B}T_{c}$ \cite{Stock1}. This intrinsic low energy scale and the strong Pauli-limited superconductivity \cite{Bianchi} in CeCoIn$_{5}$ lead to the observation of a unique behavior, the Zeeman splitting of the resonance under magnetic field \cite{Stock2,Raymond1,akbari}. In parallel to these aspects of the spin dynamics, one of the most intriguing properties of CeCoIn$_{5}$ is the occurrence, above 10.5 T, of magnetic field induced incommensurate magnetic order with $\bf{Q_{IC}}$=(0.45, 0.45, 0.5) for a magnetic field applied in the basal plane of the tetragonal structure \cite{Kenzelmann,Kenzelmann2}. Strikingly, this order disappears above the upper critical field $H_{c2}$= 11.7 T where superconductivity is suppressed, indicating a strong interplay between the superconductivity and the magnetic ordering. To date the relationship between the spin resonance and the magnetic ordering is tenuous notably due to the mismatch between their characteristic wave-vectors $\bf{Q_{AF}}$ and $\bf{Q_{IC}}$. Nonetheless a significant fact is that the extrapolation to zero energy of the lower energy mode of the Zeeman split resonance occurs in the vicinity of the onset of the Q-phase \cite{Panarin}.

In this letter, it is shown by detailed Inelastic Neutron Scattering (INS) experiments that the spin resonance is in fact incommensurate and is peaked at the same wave-vector than the propagation vector of the field induced magnetic order.
Furthermore the fluctuations associated with the resonance are found to be polarized along a unique axis, the $c$-axis and this corresponds also to the direction of the ordered magnetic moments in the sine-wave modulation of the Q-phase.
The fact that the resonance mode and the Q-phase have the same symmetry indicates that the former is a dynamical precursor of the latter and this strongly supports a scenario where the static order is realized by a condensation of the magnetic excitation. Such a mechanism falls in a broader range of condensed matter physics phenomena, associated with the so-called soft mode behavior, and ranging from lattice dynamical instability in ferroelectrics to Bose-Einstein condensation of magnons in magnetic insulators.

The experiments were performed on the recently upgraded cold neutron three axis spectrometer IN12 at ILL, Grenoble.
The initial neutron beam is provided by a double focusing pyrolitic graphite (PG) monochromator. Higher order contamination is removed before the monochromator by a velocity selector.  The spectrometer was setup in W configuration with $\alpha_{1}$-open-open collimations. 
For the experimental configuration A, a fixed $k_{f}$=1.3 $\AA^{-1}$ is used with $\alpha_{1}$=80' and a horizontally focusing PG analyzer is used.
For the experimental configuration B, a fixed $k_{f}$=1.15 $\AA^{-1}$ is used with $\alpha_{1}$=80'  and the PG analyzer is kept in flat mode (no horizontal focalization).
For the experimental configuration C, the incident neutron beam spin state is prepared by a polarizing cavity located 30 m upstream the instrument and after the velocity selector.
Guide fields, that maintain the polarization, are installed all along the neutron path including around the PG monochromator. A Mezei flipper is placed before the sample table in order to reverse the incident polarization.  At the sample position, a Helmholtz coil is used to define the direction of the polarization.
The scattered beam is analyzed by a combination of a Mezei flipper and an horizontally focusing Heusler analyzer set at fixed $k_{F}$= 1.3 $\AA^{-1}$. 
$\alpha_{1}$=open for this setup. The flipping ratio measured on the (1,1,1) and (1,1,0) Bragg peaks at $T$=1.43 K in the superconducting phase for the three polarization channels and the two flippers varies between 15 and 19. No polarization correction is applied to the data. The sample is similar to the one used in our previous studies \cite{Raymond1,Panarin} with an increased volume. It is composed of an assembly of about 80 single crystals with a total volume of about 250 mm$^{3}$. The total mosaicity of this assembly is about 1.8 degree. The sample was put in a He flow cryostat with the [1, -1, 0] axis vertical and the base temperature was 1.45 K.

\begin{figure}
\centering
\includegraphics[width=14cm]{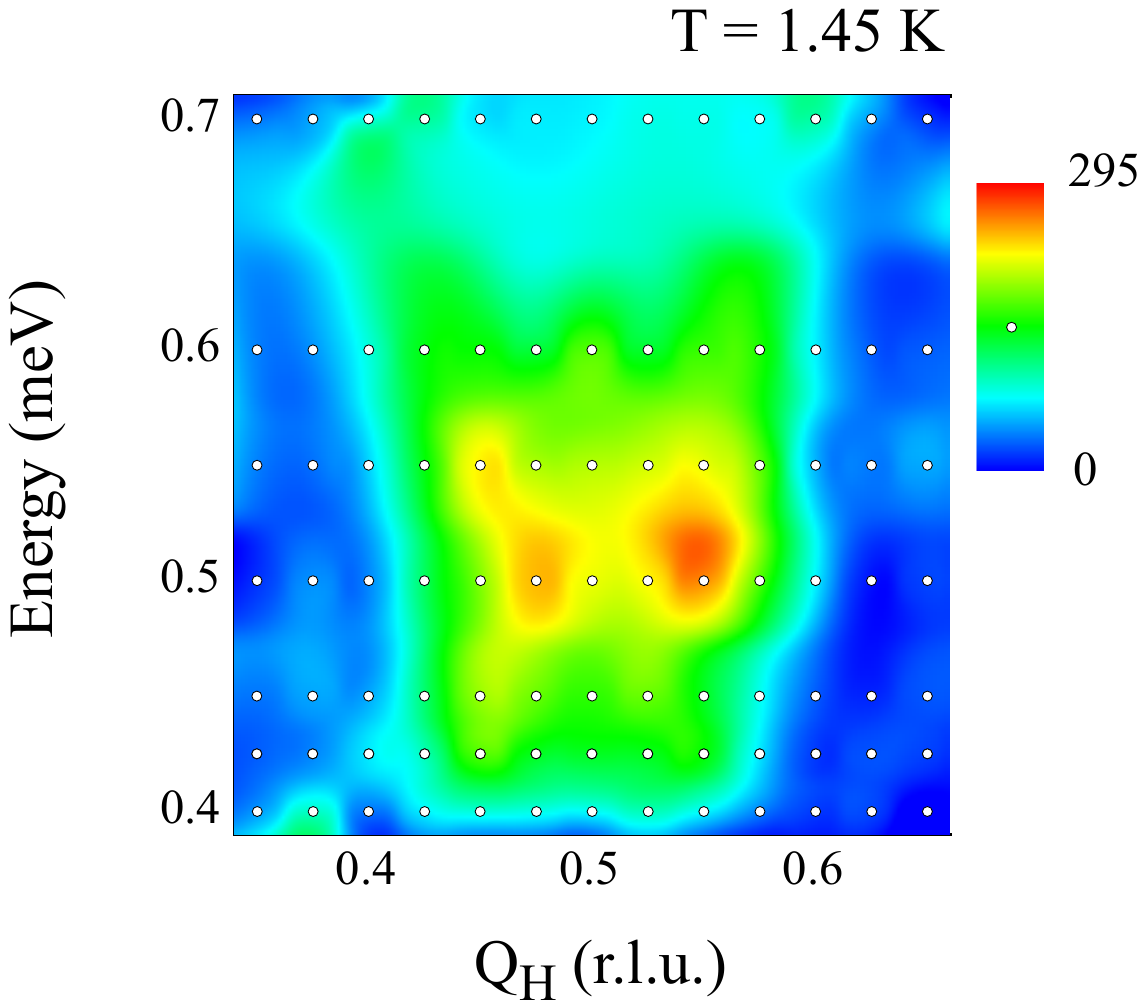}
\vspace{-12cm}
\caption{Color-coded intensity plot of the INS spectra of CeCoIn$_{5}$ as a function of $Q_{H}$ and $E$ for $\bf{Q}$=($Q_{H}$, $Q_{H}$, 0.5) at 1.45 K (Experimental configuration A). The intensity is given for a counting time of  20 minutes. The empty circles indicate positions where data were collected.}
\end{figure}

INS experiments allow to determine the spin dynamics as a function of the wave-vector $\bf{Q}$ and the energy $E$.
In the present paper, the wave-vector $\bf{Q}$ has its cartesian coordinates ($Q_{H}$, $Q_{H}$, $Q_{L}$) expressed in reciprocal lattice units (r.l.u.).
The spin resonance was precisely mapped out in the region of ($\bf{Q}$, $E$) space corresponding to its known maximum intensity with the experimental configuration A.
Focus was given on the ($Q_{H}$, $Q_{H}$, 0.5) direction that spans both $\bf{Q_{AF}}$ and $\bf{Q_{IC}}$. Figure 1 shows a color-coded intensity plot of the INS spectra collected as a function of $Q_{H}$ and $E$ at 1.45 K.
The spin resonance is clearly incommensurate in the whole measured energy range (0.4-0.7 meV) and a slight upward dispersion is observed.
A cut along ($Q_{H}$, $Q_{H}$, 0.5) at a fixed energy of 0.5 meV is shown in Figure 2a with an increased $\bf{Q}$-resolution allowing to better resolve the incommensurate peaks (Experimental configuration B).
Similar scans performed along the $c$-axis show that the signal is commensurate in this direction (Figure 2b).
A global fit of the data shown in Figure 2, using Lorentzian lineshapes, gives the position of the peaks at $\bf{Q_{AF}}$ $\pm$ ($\delta$, $\delta$, 0) with $\delta$=0.042 (2) r.l.u.. 
This value of the incommensuration determined on the inelastic correlation peaks is slightly lower than the one ($\delta_{Q}$=0.05(1)) determined on narrow elastic Bragg peaks for the Q-phase \cite{Kenzelmann2}.
Setting $\delta=\delta_{Q}$ also provides an acceptable description of the data shown in Figure 2a due to the broad nature of these correlation peaks. Therefore, it is considered that, within the experimental accuracy, the characteristic wave-vector of the maximum intensity of the spin fluctuations corresponds to the propagation vector of the ordered Q-phase. 
For completeness, it must be noticed that $\bf{Q_{IC}}$ is also the propagation vector for the magnetic ordering in slightly Nd-substituted CeCoIn$_{5}$ with $T_{N}$ $<$ $T_{c}$ \cite{RaymondNd}.
The obtained correlation length for the fluctuations at 0.5 meV are $\xi_{a}$=20(3) $\AA$ and $\xi_{c}$=12(1) $\AA$.  
The correlation along the $a$-axis is larger than the one previously reported \cite{Stock1,Panarin} due to the separation of two incommensurate peaks that were previously reported as a single peak centered at $\bf{Q_{AF}}$. 
Still the conclusion that the magnetic correlation length is smaller than the superconducting coherence length of about 40 $\AA$ applies.
Considering a local linear approximation for the upward dispersion leads to a speed of the excitation of about 4.9 meV$\AA$, which is to be related to the small exchange interactions of typically 0.5 meV characteristics of similar Ce-based heavy fermion systems \cite{Knafo,Das}.
\begin{figure}
\includegraphics[width=14cm]{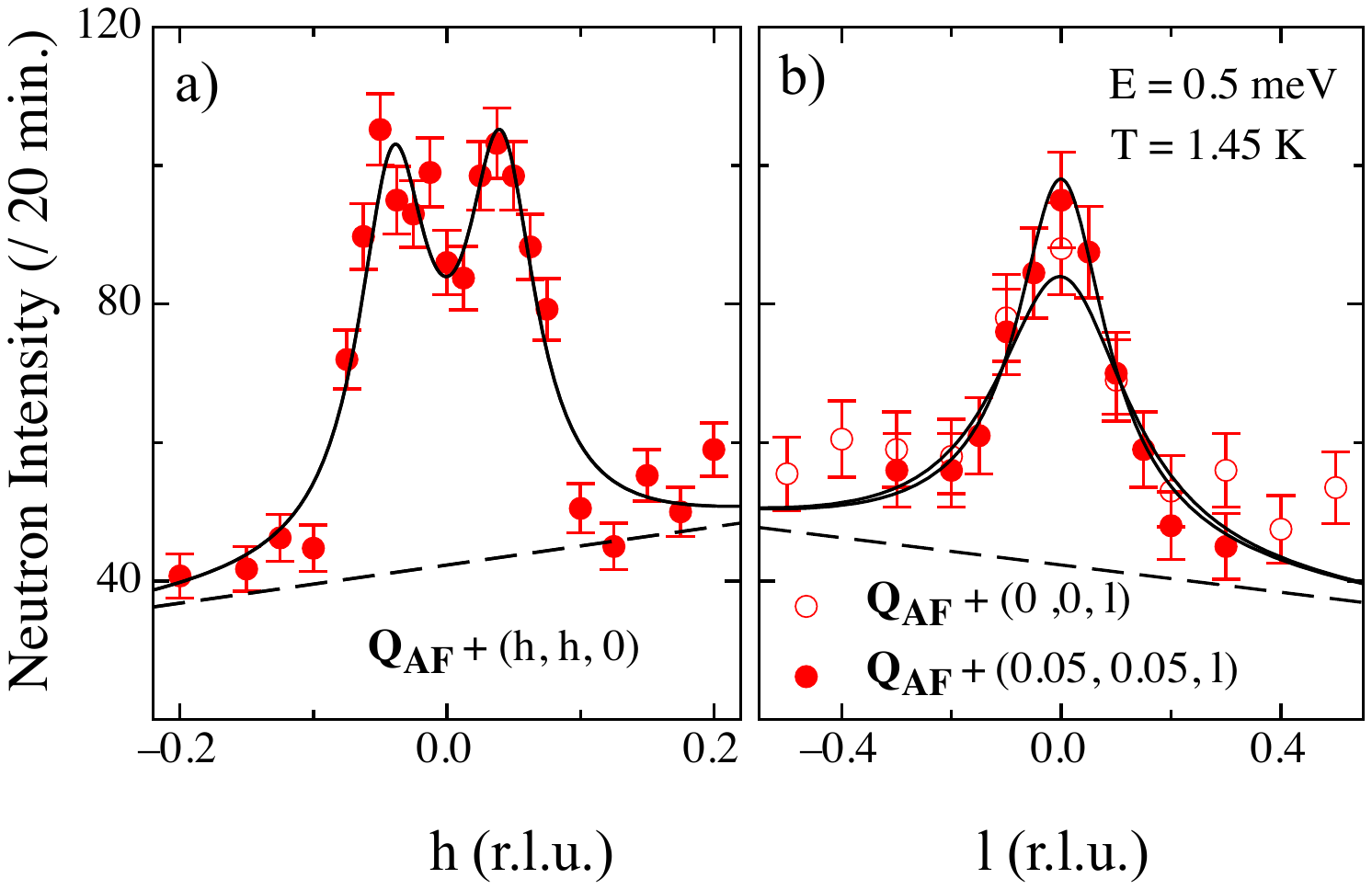}
\vspace{-5cm}
\caption{a) Constant energy scans performed at 0.5 meV for $\bf{Q}$=$\bf{Q_{AF}}$+($h$, $h$, 0) and b) at $\bf{Q}$=$\bf{Q_{AF}}$+(0, 0, $l$) and $\bf{Q}$=$\bf{Q_{AF}}$+(0.05, 0.05, $l$) (right) at $T$=1.45 K (Experimental configuration B). Solid lines correspond to a global Lorentzian fit to the data of panel a and b, with a slopping background shown as a dashed lines.}
\end{figure}

In order to determine unambiguously the spin anisotropy associated with the resonance, polarized neutron experiments were undertaken.
Because of the loss of intensity inherent to this setup, the wave-vector resolution is relaxed compared to the measurements shown above so that the splitting into incommensurate peaks is not resolved and the data are hence collected for $\bf{Q_{AF}}$ as in the previous studies.
For the polarized INS cross sections, the canonical right-handed coordinate system is used with $x$ along the scattering vector $\bf{Q}$, $y$ perpendicular to $\bf{Q}$ in the scattering plane and $z$ perpendicular to the scattering plane. In the polarized neutron cross-section, Spin Flip (SF) scattering refers to scattering process for which the final polarization is antiparallel to the initial one. The measured double differential neutron cross-section for SF scattering and polarization along the axis $\alpha$, is written $\sigma^{SF}_{\alpha}$ with:
\begin{eqnarray}
\sigma_{x}^{SF} &\propto  & BG_{SF}+1.39M^{a}_{Q,\omega}+0.92M^{c}_{Q,\omega}\\
\sigma_{y}^{SF}  & \propto & BG_{SF}+M^{a}_{Q,\omega}\\
\sigma_{z}^{SF} & \propto & BG_{SF}+0.39M^{a}_{Q,\omega}+0.92M^{c}_{Q,\omega}
\end{eqnarray}
where  $BG_{SF}$ is the background for SF channel that includes for convenience the contributions from nuclear spin scattering and $M^{\beta}_{Q,\omega}=\frac{1}{2\pi}\int <M(-Q)^{\beta}_{\perp}(0)M(Q)^{\beta}_{\perp}(t)>e^{-i\omega t}dt$ where $M(Q)^{\beta}_{\perp}(t)$ is the $\beta$ component in the sample frame ($\beta$=$a$, $c$) of the Fourier component of the sample magnetization perpendicular to $\bf{Q}$ and $<..>$ is the quantum statistical expectation value. The conversion between the cartesian $x$, $y$, $z$ coordinates and the crystal axes [1, 0, 0], [0, 1, 0] and [0, 0, 1] is made considering the angle of 23$^{\circ}$ between the [1, 1, 1] and the [1, 1, 0] directions and making the hypothesis of isotropic fluctuation in the plane (equivalence between [1,0,0], [0,1,0] and [1,1,0] directions).

\begin{figure}
\includegraphics[width=12cm]{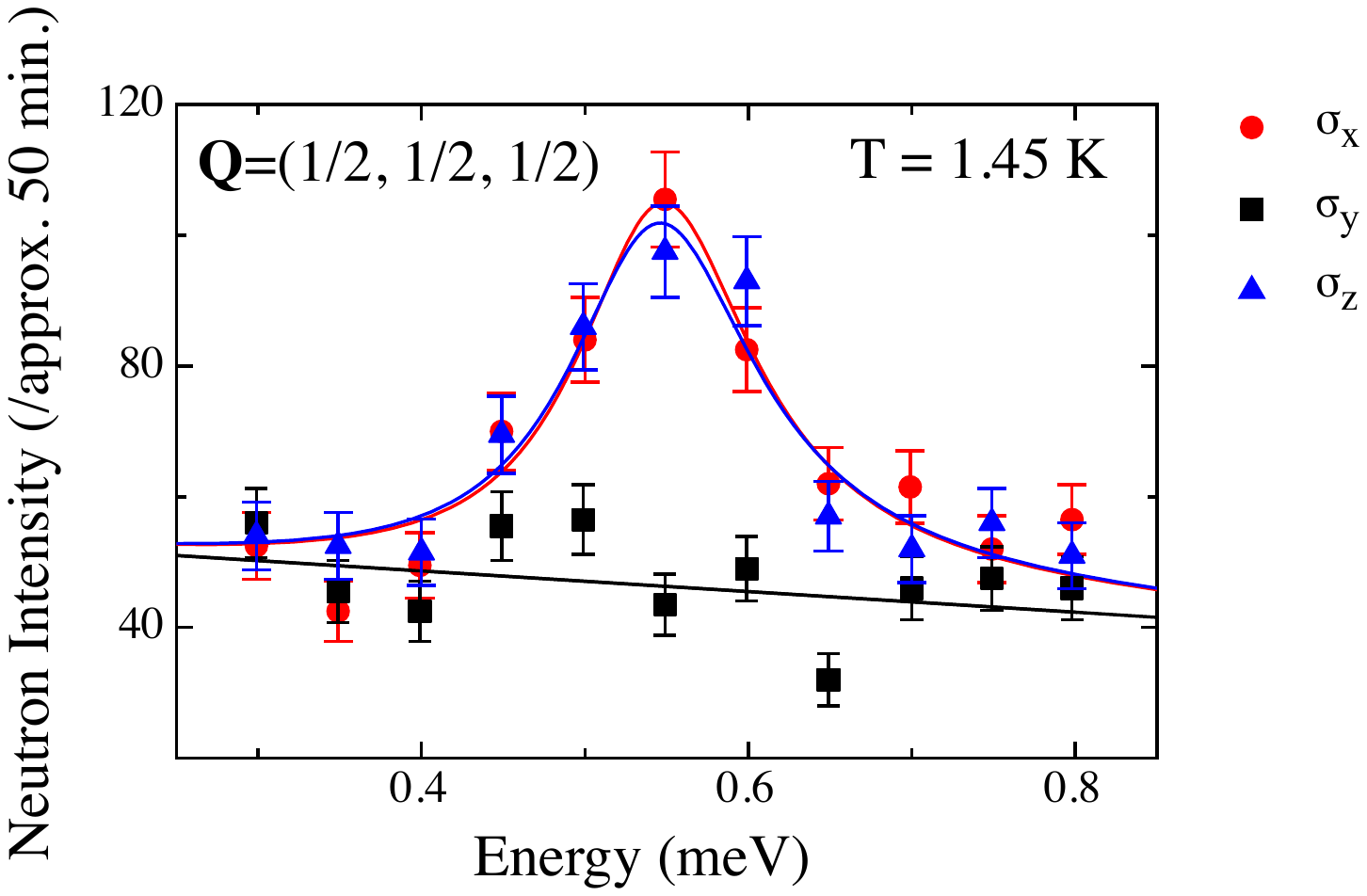}
\vspace{-11cm}
\caption{Polarized inelastic neutron scattering spectra taken at $\bf{Q_{AF}}$=(0.5, 0.5, 0.5) corresponding to Spin Flip scattering for neutron polarization along $x$, $y$, $z$ (Experimental configuration C). Lines are fit to the data using a $\omega$-Lorentzian lineshape as in Ref.\cite{Panarin} and a sloppping background.}
\end{figure}
Figure 3 shows an energy spectra measured at 1.45 K at $\bf{Q_{AF}}$ for SF scattering using the Mezei flipper located after the sample with the polarization along $x$, $y$ and $z$. The spin resonance excitation is observed for the polarization along the $x$ and $z$ axis with the same intensity while the scattering for the polarization along the $y$ axis is structureless and corresponds to the background. Fits of the data are performed using an $\omega$-Lorentzian as in a previous work \cite{Panarin} and consistently, the energy of the resonance is found to be $\Omega_{res}$=0.54(1) meV and the linewidth  $\Gamma$=0.07(1) meV, which is about twice the resolution.
An inspection of these data in view of Eq.(1)-(3) indicates that $M^{a}_{Q,\omega}$ is zero and $M^{c}_{Q,\omega}$ is non vanishing.  Hence the fluctuations associated with the resonance peak are polarized along the $c$-axis of the tetragonal structure without any in-plane contribution. The $c$-axis is also the direction of the amplitude modulated ordered magnetic moments in the magnetic field induced Q-phase. 
The exact same conclusions are reached for the data (not shown) collected in the Non Spin Flip channel.

In this paper, it is established by INS that (i) the spin resonance of CeCoIn$_{5}$ is incommensurate and located at $\bf{Q_{IC}}$ (ii) the corresponding fluctuations are polarized solely along the $c$-axis.
These two characteristic features are shared with the Q-phase : it is an incommensurate sine-wave modulated structure of propagation vector $\bf{Q_{IC}}$ and the magnetic moments are ordered along the $c$-axis.
The fact that the dynamical mode at $H$=0 T and the field induced static order share to the same symmetry underlines the fact that the resonance is a dynamical precursor of the Q-phase.
This, together with the known softening of the lowest energy mode of the Zeeman split resonance \cite{Stock2,Raymond1,Panarin}, strongly supports the theoretical scenario where the magnetic ordering is obtained by a field induced condensation of the resonance. Such a scenario was theoretically put forward in a microscopic model by Michal and Mineev \cite{Michal} but, up to the present study, there was (i) an important mismatch between the characteristic wave-vector of the spin dynamics and the one of the magnetic order and (ii) ambiguities about the spin fluctuation anisotropy \cite{note}.

Field induced or field-enhanced magnetic ordering out of a superconducting phase is well-known for cuprates \cite{Lake}.
The soft-mode scenario was also considered on phenomenological ground for La$_{1.855}$Sr$_{0.145}$CuO$_{4}$ where the low energy spin-gap decreases with magnetic field and collapses when antiferromagnetism appears \cite{Chang}. 
A common view for cuprates is that the magnetic field reveals the underlying ground state of the normal (non superconducting) phase in a generic quantum critical phase diagram \cite{Demler}.
This view is in stark contrast with the situation of CeCoIn$_{5}$ where the cooperative nature of the interaction between magnetism and $d$-wave superconductivity is  already obvious from the disappearance of magnetic ordering at $H_{c2}$ and their tight microscopic relation is furthermore highlighted by the pinning of $\bf{Q_{IC}}$ to the superconducting gap nodal direction \cite{Kenzelmann2} and the switching of magneto-superconducting domains \cite{Gerber}. 

The specificities of a 4$f$-electron system like CeCoIn$_{5}$ opens the way to new aspects absent for the $d$-electron based unconventional superconductors and this can help understanding more generally the nature of the spin resonance and consequently the unconventional superconducting state itself.
However, the nature of the resonance peak is not completely settled and a spin-exciton \cite{Eremin}, the most common model \cite{Eschrig}, or a magnon-like excitation \cite{Chubukov} are proposed for CeCoIn$_{5}$. 
Four main characteristics of the spin resonance of CeCoIn$_{5}$ are evidenced in the present work, namely the incommensurability, the upward dispersion, the modest exchange and the uniaxial fluctuations.
The first three characteristics are shared with the resonance mode of CeCu$_{2}$Si$_{2}$ \cite{Stockert}, suggesting possible distinctive features of  the excitation spectrum of paramagnetic heavy fermion superconductors. These specificities, reminiscent of "heavy-fermion physics", underline the role of the Fermi surface topology and the RKKY nature of the magnetic interactions. 
While the Fermi surface is composed of multiple quasi-two-dimensional sheets \cite{Settai}, a single band is pointed out to be the relevant one when wave-vectors in the vicinity of  $\bf{Q_{AF}}$ are involved \cite{Allan,Zhou}.
However to date and despite their increasing accuracy, experimental and theoretical band structure determinations of CeCoIn$_{5}$ do not reveal the fine details of the Fermi surface topology that define the vector $\bf{Q_{IC}}$ \cite{Nomoto}.
Because of this lack of knowledge, it seems premature to draw any conclusion from the observed upward dispersion that is predicted in the magnon-like model \cite{Chubukov} while a downward dispersion is predicted for the spin-exciton model \cite{Eremin}. Indeed both models consider a commensurate resonance located at $\bf{Q_{AF}}$. Our finding of incommensurate resonance therefore asks to revisit these models and their corresponding dispersion. Finally, it must be underlined that the Ising nature of the fluctuations is unique among resonance excitations in unconventional superconductors and arises from the crystal field anisotropy governing such 4$f$-electron systems. In contrast, for $d$-electron based unconventional superconductors, where the origin of anisotropy is the spin-orbit coupling, polarized INS data reveal an overall modest splitting in energy of in-plane and out-of-plane components of the fluctuations for Fe-based superconductors \cite{Inosov} or cuprates \cite{Headings}.

The new relationship between the spin resonance excitation and the ordered Q-phase of CeCoIn$_{5}$ that evidences their common symmetry strongly supports the scenario where the Q-phase is obtained by a spin excitation condensation under magnetic field. This underlines the application of the generic concept, transverse to condensed matter physics, of soft mode for a complex system where magnetism and superconductivity are strongly interconnected.

\bigskip

We acknowledge L.P. Regnault, Y. Sidis and V.P. Mineev for illuminating discussions and K. Mony and B. Vettard for their technical support during sample preparation and neutron scattering experiment.

\end{document}